\documentclass[%
 reprint,
 superscriptaddress,
 amsmath,amssymb,
 aip,
 jcp,
floatfix,
]{revtex4-2}

\usepackage{color,graphics}
\usepackage{graphicx}
\usepackage[sort&compress]{natbib}
\usepackage{hhline}
\usepackage{soul}
\usepackage{url}
\usepackage{hyperref}

\usepackage{graphicx}
\usepackage{dcolumn}
\usepackage{bm}

\usepackage[utf8]{inputenc}
\usepackage[T1]{fontenc}
\usepackage{mathptmx}
\usepackage{newtxtext,newtxmath} 
\usepackage{etoolbox}

\usepackage[usenames,dvipsnames]{xcolor}

\definecolor{OGreen}{rgb}{0,0.6,0}

\makeatletter
\def\@email#1#2{%
 \endgroup
 \patchcmd{\titleblock@produce}
  {\frontmatter@RRAPformat}
  {\frontmatter@RRAPformat{\produce@RRAP{*#1\href{mailto:#2}{#2}}}\frontmatter@RRAPformat}
  {}{}
}%
\makeatother
\begin{document}

\preprint{
}

\title[]{
Donnan equilibrium in charged slit-pores from a hybrid nonequilibrium Molecular Dynamics / Monte Carlo method with ions and solvent exchange}
\author{Jeongmin Kim}
\affiliation{Department of Energy Engineering, Korea Institute of Energy Technology (KENTECH), Naju 58330, Republic of Korea}
\author{Benjamin Rotenberg}
\email{benjamin.rotenberg@sorbonne-universite.fr}
\affiliation{Sorbonne Universit\'{e}, CNRS, Physico-chimie des \'{E}lectrolytes et Nanosystem\`{e}s Interfaciaux, PHENIX, F-75005 Paris, France}
\affiliation{R\'eseau sur le Stockage Electrochimique de l'Energie (RS2E), FR CNRS 3459, 80039 Amiens Cedex, France}

\date{\today}

\begin{abstract}
Ion partitioning between different compartments (\emph{e.g.} a porous material and a bulk solution reservoir), known as Donnan equilibrium, plays a fundamental role in various contexts such as energy, environment, or water treatment. The linearized Poisson-Boltzmann (PB) equation, capturing the thermal motion of the ions with mean-field electrostatic interactions, is practically useful to understand and predict ion partitioning, despite its limited applicability to conditions of low salt concentrations and surface charge densities. Here, we investigate the Donnan equilibrium of coarse-grained dilute electrolytes confined in charged slit-pores in equilibrium with a reservoir of ions and solvent. We introduce and use an extension to confined systems of a recently developed hybrid nonequilibrium molecular dynamics / grand canonical Monte Carlo simulation method ("H4D"), which enhances the efficiency of solvent and ion-pair exchange via a fourth spatial dimension. We show that the validity range of linearized PB theory to predict the Donnan equilibrium of dilute electrolytes can be extended to highly charged pores, by simply considering \textit{renormalized} surface charge densities. We compare with simulations of implicit solvent models of electrolytes and show that in the low salt concentrations and thin electric double layer limit considered here, an explicit solvent has a limited effect on the Donnan equilibrium and that the main limitations of the analytical predictions are not due to the breakdown of the mean-field description, but rather to the charge renormalization approximation, because it only focuses on the behavior far from the surfaces.
\end{abstract}

\maketitle

\section{Introduction}
The composition of an electrolyte near a charged surface or in a charged porous material differs from that of a bulk solution reservoir with which it is in equilibrium. This ion partitioning between different compartments, known as Donnan equilibrium~\cite{Donnan1911theory}, plays an important role in biology~\cite{nelson2020biological} or in the environment~\cite{jardat2009salt,barry2021advanced,hsiao2022salt,ilgen2023adsorption}; understanding and predicting it is necessary for various applications \emph{e.g.} water treatment~\cite{briskot2022modeling}, membrane technology~\cite{galama2013validity,siria2017new,gao2022increased,aydogan2022Donnan}, energy~\cite{kamcev2016charged}, or electrocatalysis~\cite{shin2022importance,xu2024cation}. The classical approach to this Donnan equilibrium is to consider the equality of the electrochemical potentials of the various species in the reservoir and in the "phase" or region of interest. This results in the celebrated quadratic relation between the salt concentrations in the reservoir and the confined medium, and a so-called Donnan potential difference between the two regions~\cite{Donnan1911theory}. This Donnan model can be refined to include various interactions between the ions or with the medium~\cite{biesheuvel2011theory,misra2019theory,chen2022Donnan,aydogan2022Donnan,gao2022increased}.

This simple approach neglects the inhomogeneities in the distribution of ions and the corresponding electrostatic potential inside the pores. In fact, the balance between electrostatic interactions (energy) and the thermal motion of ions (entropy) results near charged walls in electric double-layers (EDL), with an enrichment of counterions and a depletion of co-ions with respect to the bulk~\cite{parsons1990electrical,bard2001fundamentals,markovich2021charged,herrero2021Poisson}. Poisson-Boltzmann (PB) theory can capture the main features of such EDL by considering point charges in an implicit solvent, and treating electrostatic interactions at the mean-field level. For sufficiently small salt concentrations and surface charge densities, PB theory can further be linearized, resulting in Debye-H\"uckel (DH) theory, which predicts an exponential decay of density and potential profiles as a function of the distance from the wall, over the so-called Debye length $\lambda_D$. The classical Donnan approach typically applies in the limit where this length is large compared to the pore size, so that spatial variations can be neglected. PB or DH theory can be used in simple geometries to predict ion partioning as a function of pore size, surface charge density and salt concentration in the reservoir. However its applicability is usually limited to conditions of low salt concentrations and surface charge densities. For larger surface charge densities, the large concentration of counterions (overestimated by PB theory, which neglects in particular the excluded volume interactions due to the finite size of the ions) is usually lumped into a (Stern) layer of condensed ions, resulting in the so-called Gouy-Chapman-Stern theory of the interface~\cite{parsons1990electrical,bard2001fundamentals}. 

Despite its great merits, this simplified picture of the EDL fails to account for microscopic features which may play an important role at the interface. For example, the molecular nature of the solvent results in packing of hydration effects known to influence the distribution of ions at the surface~\cite{park2006hydration,ricci2014water,brown2016effect,brown2016determination,hartkamp2018measuring,ma2021stern,lee2021ion,hunger2022nature,gao2022increased}, or the force between atomically flat mica surfaces in Surface Force Balance experiments (SFB)~\cite{smith2016electrostatic,lee_underscreening_2017,elliott_known-unknowns_2024}. Efforts have been made to modify PB theory to account for effects such as the finite size of ions~\cite{torrie1980electrical,torrie1982electrical,outhwaite1980theory,lopez2011Poisson}, ion-surface interactions~\cite{bostrom2001specific,joly2004hydrodynamics,joly2006liquid,aydogan2022Donnan}, electrostatic correlations~\cite{netz2000beyond}, solvent polarization~\cite{hedley2023dramatic,schlaich2024water}, or surface charge regulation~\cite{alexander1984charge,belloni1998ionic,aubouy2003effective,kreer2006nonlinear,trefalt2016charge,acharya2020charge,schlaich2023renormalized,brito2023effective}.

In order to go beyond these theories, Grand-canonical (GC) Monte Carlo (MC) simulations have been used to investigate the Donnan equilibrium of confined electrolytes in equilibrium with a reservoir~\cite{valeau1980primitive,smit1995grand,stern2007molecular,barr2012grand,michael2020hybrid}. However, while GC simulations of electrolytes described as ions in an implicit solvent can be performed with sufficiently large acceptance rates of trial insertion/deletion MC moves to allow for a proper sampling of configurations, such simulations with an explicit solvent are much more challenging, not only due to the larger computational cost of interactions, but also to the very low acceptance rate of such moves. Several advanced MC methods to enhance exchange efficiency have been developed, including a cavity-bias method~\cite{mezei1980cavity}, a configuration-biased method~\cite{shelley1994configuration,shelley1995configuration,smit1995grand}, a Boltzmann-bias method~\cite{garberoglio2008Boltzmann}, a continuous fraction component method~\cite{shi2007continuous,moucka2015electrolyte}, identity exchange methods~\cite{panagiotopoulos1989exact,soroush2018molecular,fathizadeh2018mixed}, the adaptive resolution method~\cite{praprotnik2005adaptive,baptista2021density}, and hybrid methods~\cite{duane1987hybrid,mehlig1992hybrid,boinepalli2003grand,stern2007molecular,chen2014efficient,radak2017constant,ross2018biomolecular,prokhorenko2018large,nilmeier2011nonequilibrium,belloni2019non,kim2023jcp}. Recently, Belloni introduced a hybrid MC/nonequilibrium molecular dynamics (MD) method~\cite{belloni2019non}, called "H4D", with trial moves during which the system adjusts to the gradual addition/deletion of particles via a non-physical additional dimension. This results in a significant increase in the acceptance rate of such moves, as demonstrated \textit{i.e.} for both ion-pairs and solvent exchange for bulk aqueous NaCl electrolytes~\cite{belloni2019non,kim2023jcp}.

Here, we extend the H4D method to the case of confined systems and use it to investigate the Donnan equilibrium of coarse-grained dilute electrolytes in charged slit-pores in equilibrium with a reservoir of ions and solvent. We show that the validity range of linearized PB theory to predict the Donnan equilibrium of dilute electrolytes can be extended to highly charged pores, by considering renormalized surface charge densities~\cite{alexander1984charge,belloni1998ionic,aubouy2003effective,kreer2006nonlinear,schlaich2023renormalized,brito2023effective}, which can be computed analytically by comparing the full and linearized PB equations, instead of the bare ones. By comparing with simulations of implicit solvent models of electrolytes, we find that for the small salt concentration considered here, an explicit solvent introduces oscillations in the ionic density profiles, but has a limited effect on the excess salt concentration inside the pore. In the low concentration and thin electric double-layer limit considered here, the main limitations of the analytical predictions are not due to the breakdown of the mean-field description, but rather to the charge renormalization approximation, which is not sufficient to capture the overall excess ion density because it focuses on the behavior far from the walls. 

The rest of the manuscript is organized as follows. Section~\ref{sec:donnan} discusses the classical description of the Donnan equilibrium as well as its prediction from Poisson-Boltzmann theory and the surface charge renormalization approach. Section~\ref{sec:methods} summarizes the basics of the H4D method for grand-canonical molecular dynamics (GCMD) simulations introduced in Refs.~\cite{belloni2019non,kim2023jcp} and extends it to confined electrolytes. Section~\ref{sec:methods} also presents the model system for which we studied the Donnan equilibrium: Lennard-Jones model electrolytes confined in a charged slit-like pore. Section~\ref{sec:results} then discusses the results of GCMD simulations and their comparison with the linearized PB theory, including the surface charge renormalization approach. The main conclusions are summarized in Section~\ref{sec:conclusion}

\section{Donnan equilibrium of electrolytes between a pore and a reservoir}
\label{sec:donnan}

\begin {figure}[htbp]
\includegraphics [width=3in] {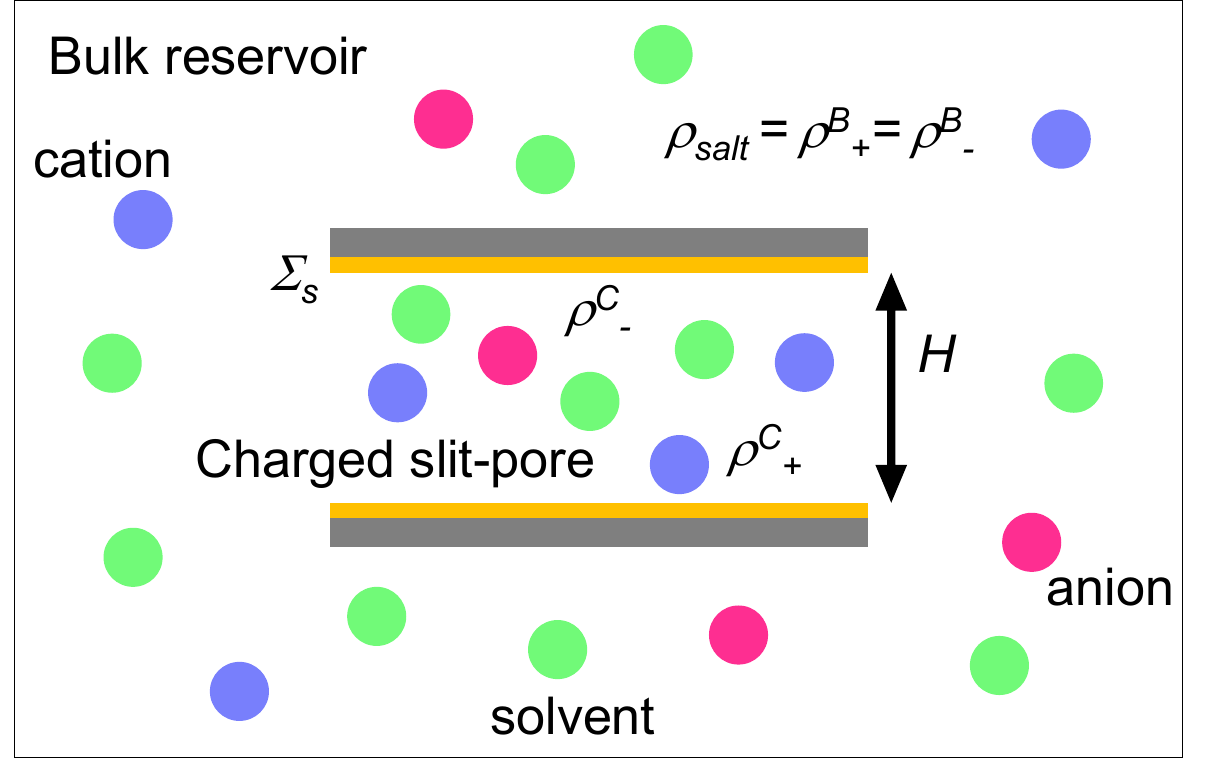}
\caption{Donnan equilibrium for a charged slit-pore in contact with a bulk reservoir. Here, $H$ and $\Sigma_s$ indicate the width and the surface charge density of the pore, respectively. The ion concentrations in the bulk and inside the pore are indicated by $\rho^B_i$ and $\rho^C_i$ with $i\in\{+,-\}$, respectively. See Section~\ref{sec:donnan:donnan} for details.
}
\label{fig:schemdonnan}
\end{figure}

As mentioned above, the composition of an electrolyte near a charged surface or in a charged porous material differs from that of a bulk solution reservoir with which it is in equilibrium (See Fig.~\ref{fig:schemdonnan}). Here, we first recall the classical Donnan description of this equilibrium (Section~\ref{sec:donnan:donnan}) before turning to the predictions obtained for a charged slit pore using the Poisson-Boltzmann equation and its linearized version (Section~\ref{sec:donnan:pb}). We finally introduce an effective surface charge density of the pore walls by comparing the full and linearized Poisson-Boltzmann equations, according to the charge renormalization approach: a fraction of the counterions is included as part of the surface charge, thereby reducing the latter sufficiently for the linearization to be a good approximation far from the surface~\cite{attard1995ion,belloni1998ionic} (Section~\ref{sec:donnan:effectivecharge}).

\subsection{Classical Donnan approach}
\label{sec:donnan:donnan}

At equilibrium, the electrochemical potential of the various species is the same inside the pore of interest and in the reservoir with which the solution exchanges particles. For ionic species $i\in \{+,-\}$ with charges $q_\pm=\pm e$ with $e$ the elementary charge, the equality of electrochemical potential between the bulk reservoir $B$, with concentrations $\rho_i^B$ and potential $\Psi^{B}$, and the confined region $C$, with position-dependent concentrations and potential $\rho^{C}_i({\bf r})$ and $\Psi^{C}({\bf r})$, reads:
\begin{equation}
\label{eq:donnan}
\begin{split}
\ln(\gamma^{B}_i\rho^{B}_i/\rho_\text{ref})+\beta q_i\Psi^{B}=\\
\ln[\gamma^{C}_i({\bf r})\rho^{C}_i({\bf r})/\rho_\text{ref}]+\beta q_i\Psi^{C}({\bf r}),
\end{split}
\end{equation}
where $\gamma_i$ is the activity coefficient of species $i$, $\rho_\text{ref}$ is a reference concentration and $\beta=1/k_BT$ with $k_B$ the Boltzmann constant and $T$ the temperature. 
Then, the ion concentrations in the confined region is:
\begin{equation}\label{boltz_dist}
\begin{split}
\rho^{C}_i({\bf r})=\rho^{B}_i\frac{\gamma^{B}_i}{\gamma^{C}_i({\bf r})}\exp[-\beta q_i(\Psi^{C}({\bf r})-\Psi^{B})] \; .
\end{split}
\end{equation}
This can be further simplified by assuming an ideal behavior of the electrolyte (activities $\gamma_i=1$) and no spatial dependence of the potential and of the ion concentrations inside the pore. Assuming without loss of generality that the surface is negatively charged, with a corresponding excess of cations inside the pore with concentration $\rho^{C}_{+,ex}$ counterbalancing the surface charge, one obtains the well-known quadratic equation for the salt concentration $\rho^{C}_{\text{salt}}$ in the pore:
\begin{equation}\label{donnan_ion}
\begin{split}
(\rho_{\text{salt}})^2 
= \rho^{C}_{+}\rho^{C}_{-}=(\rho^{C}_{\text{salt}}+\rho^{C}_{+,\text{ex}})\rho^{C}_{\text{salt}} \;,
\end{split}
\end{equation}
where $\rho_{\text{salt}}=\rho^{B}_{\pm}=\rho^B_{\text{salt}}$, the salt concentration in the electrically neutral reservoir. Inside a charged pore, the co-ion (resp. counterion) density is expected to decrease (resp. increase) with respect to the bulk salt density, as a result of the balance between electrostatic and osmotic pressures~\cite{barrat2003basic}. As in Equation~\ref{boltz_dist}, the electric potential determines the ion distributions in a confined region. 

\subsection{Linearized Poisson-Boltzmann (Debye-H\"uckel) theory}
\label{sec:donnan:pb}

For a 1:1 electrolyte in a symmetric charged slit-pore, with two walls with surface charge density $\Sigma_s$ located at $z=\pm H/2$, one obtains the one-dimensional (1D) electrostatic potential profile $\Psi(z)$ by solving the following 1D Poisson-Boltzmann equation (taking the electrostatic potential in the reservoir as reference $\Psi_B=0$)~\cite{barrat2003basic}: 
\begin{equation}\label{eq:pb}
\frac{d^2}{dz^{2}}\Psi(z)=-\frac{\rho_q(z)}{\varepsilon_0\varepsilon_s}
=\frac{\kappa^{2}_D}{\beta e}\sinh[\beta e\Psi(z)]
\end{equation}
where $\rho_q(z)= e\left[\rho^{C}_+(z)-\rho^{C}_-(z)\right]$ is the local charge density, $\varepsilon_s$ is the solvent dielectric constant, and $\kappa_{D}=\lambda^{-1}_{D}$, with 
\begin{equation}
    \label{eq:debye}
    \lambda_D=\sqrt{\frac{\varepsilon_0\varepsilon_s}{2\beta\rho_{\text{salt}}e^2}}
\end{equation} the Debye screening length, with $\rho_{\text{salt}}$ the salt concentration in the electroneutral reservoir which also sets the potential reference. For sufficiently low surface charge densities, potential variations across the slit-pores are small ($|e\beta\Psi(z)|\ll1$) and Eq.~\ref{eq:pb} can be linearized, resulting in:
\begin{equation}\label{eq:dhb2}
\Psi(z)=\Psi_0\frac{\cosh(\kappa_{D}z)}{\sinh(\kappa_{D}H/2)},
\end{equation}
where $\Psi_0=\Sigma_s/\varepsilon_s\kappa_D$ is the potential at the walls. Accordingly, the mean ion concentrations ($\bar{\rho}_{+}$ and $\bar{\rho}_{-}$) inside the negatively charged pore can be computed from the linearized Boltzmann distributions 
$\rho^{C}_\pm(z)=\rho_{\text{salt}}\left[1 \mp \beta e \Psi(z)\right]$:
\begin{equation}\label{lpb+}
\begin{split}
\bar{\rho}_+(H)&=\frac{1}{H}\int_{-H/2}^{{H}/2}dz~\rho^C_+(z)=\rho_{\text{salt}}+\frac{|\Sigma_s|}{eH}. 
\end{split}
\end{equation}
and
\begin{equation}\label{lpb-}
\bar{\rho}_-(H)=\frac{1}{H}\int_{-H/2}^{{H}/2}dz~\rho^{C}_-(z)=\rho_{\text{salt}}-\frac{|\Sigma_s|}{eH}.
\end{equation}
The mean ion densities deviate from their bulk value, depending on the pore size $H$, with an enrichment in counterions and depletion of co-ions, as expected. Furthermore, these deviations are equal and opposite, so that the total ion density ($2\rho_{\text{salt}}$) is independent of the pore size and the excess ion density vanishes ($\Delta\rho_{\text{ex}}(H)=\bar{\rho}_+(H)+\bar{\rho}_-(H)-2\rho_{\text{salt}}=0$).

\subsection{Effective surface charge density} 
\label{sec:donnan:effectivecharge}

\begin {figure}[htbp]
\includegraphics [width=3in] {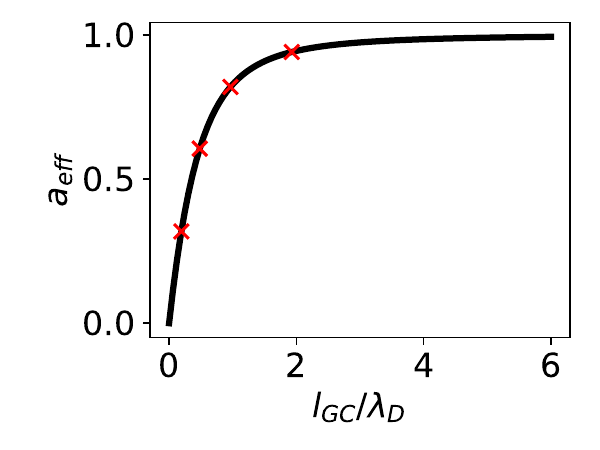}
\caption{Charge renormalization parameter $a_{\text{eff}}$ (see Eq.~\ref{eq:aeff}) characterizing the effective surface charge density, as a function of the ratio between Gouy-Chapman length $l_{GC}$ and Debye screening length $\lambda_D$. Four red markers indicates the conditions at which simulations in this study were conducted, with a fixed $\lambda_D$ corresponding to $\rho_{\text{salt}}=0.0144$ in reduced LJ units (see Section~\ref{sec:methods}), while $l_{GC}$ varies according to the surface charge density.
}
\label{fig:aeff}
\end{figure}

When the distance between the confining walls is much larger than the thickness of the EDL ($\lambda_D\ll H$), the so-called thin-EDL or single-wall approximation works well, resulting in the following analytical solution of the non-linear Eq.~\ref{eq:pb} near each wall:
\begin{equation}
\label{eq:dhb}
\Psi(z_d)= \frac{4}{\beta e}\tanh^{-1}[\gamma \exp(-\kappa_{D}z_d)]
\end{equation}
with $z_d$ the distance from the wall and the dimensionless parameter $\gamma\in\left.\right[0,1\left]\right.$, determined by the boundary condition (fixed surface charge density $\Sigma_s$), is 
\begin{equation}
    \label{eq:gamma}
    \gamma=-l_{GC}\kappa_{D}+\sqrt{(l_{GC}\kappa_{D})^2+1} 
\end{equation}
with the Gouy-Chapman length 
\begin{equation}
    \label{eq:gc}
    l_{GC}=\frac{2\varepsilon_0\varepsilon_s}{\beta e|\Sigma_s|}. 
\end{equation}
At large distances from the charged surface, the electric potential in Eq.~\ref{eq:dhb} decays as $\frac{4\gamma}{\beta e}\exp(-\kappa_{D}z_d)$. Comparison with the solution of the linearized solution Eq.~\ref{eq:dhb2} then allows to define an effective surface charge density such that the two solutions coincide at large distances~\cite{attard1995ion,buyukdagli2012electrostatic,buyukdagli2016beyond}, as $\Sigma_{s,\text{eff}}=a_{\text{eff}}\Sigma_s$, with:
\begin{equation}\label{eq:aeff}
a_{\text{eff}}=2\gamma l_{GC}\kappa_D .
\end{equation}
This effective surface charge accounts for the screening of the electric field by the ions close to the surface, up to the point where it has sufficiently decayed so that the linearization becomes accurate. This is closely related to the concept of charge renormalization \cite{alexander1984charge,belloni1998ionic,aubouy2003effective,kreer2006nonlinear,schlaich2023renormalized,brito2023effective}, widely adopted in the colloid community.
As shown in Fig.~\ref{fig:aeff}, $a_{\text{eff}}$ increases from 0 for highly charged surfaces ($l_{GC}\lesssim\lambda_D$), reflecting the fact that the strong attraction of counterions towards the wall screens most of the electric field at the surface, to 1 for lightly charged surfaces ($l_{GC}\gtrsim\lambda_D$). We note that the expression of $a_{\text{eff}}$ in Eq.~\ref{eq:aeff} is still at the mean field level; the contribution of ionic correlations can further decrease $a_{\text{eff}}$, particularly at high $\Sigma_s$.~\cite{buyukdagli2012electrostatic,buyukdagli2016beyond}

Assuming that the co-ions (here anions) are sufficiently far from the surface for the linearization to hold even in the presence of non-linear behaviour close to the surface, \emph{i.e.} that only counterions (here cations) are affected by such non-linearity, we can replace the non-linear potential profile by its linearized counterpart for an effective surface charge density $a_{\text{eff}}\Sigma_s$ in the $\rho^{C}_-(z)=\rho_{\text{salt}}\left[1 + \beta e \Psi(z)\right]$. This leads to the average anion concentration in the slit-pore:
    \begin{equation}\label{mlpb-}
      \bar{\rho}_-(H)=\frac{1}{H}\int_{-H/2}^{H/2}dz~\rho^{C}_-(z)=\rho_{\text{salt}}-a_{\text{eff}}\frac{|\Sigma_s|}{eH} \; .
    \end{equation}
Further assuming the overall electroneutrality of the system, \textit{i.e.} that the cations compensate both the bare surface charge density and the presence of the anions in equilibrium with the reservoir, yields the average cation concentration:
    \begin{equation}\label{mlpb+}
    \bar{\rho}_+(H)=\frac{1}{H}\int_{-H/2}^{H/2}dz~\rho^C_+(z)=\rho_{\text{salt}}+(2-a_{\text{eff}})\frac{|\Sigma_s|}{eH} \; .
    \end{equation}
    These results reduce to Eqs.~\ref{lpb+}-~\ref{lpb-} in the limit $l_{GC}/\lambda_D\to\infty$, where $a_{\text{eff}}=1$. The salt concentration in the pore ($\rho^{C}_{\text{salt}}=\rho^{C}_{-}$) differs from that in the reservoir ($\rho_{\text{salt}}$) and this Donnan equilibrium is quantified by the excess average ion density, which at this level of description reads:
    \begin{equation}\label{eq:excess_mean_eff}
    \Delta\rho_{\text{ex}}=\bar{\rho}_+(H)+\bar{\rho}_-(H)-2\rho_{\text{salt}}=2(1-a_{\text{eff}})\frac{|\Sigma_s|}{eH}.
    \end{equation}
    We recall that the above results involving the renormalized surface charge density assume sufficiently large pores, \textit{i.e.} $H\gg l_{GC}, \lambda_D$.

\section{Model systems and simulation methods}\label{sec:methods}

Here, we introduce the Lennard-Jones (LJ) electrolytes confined in a charged slit-pore studied in this work (Section~\ref{sec:methods:model}), and present the hybrid nonequilibrium MD / Monte Carlo algorithm with particle exchange via a fourth spatial dimension, which we extend to the case of confined electrolytes (Section~\ref{sec:methods:h4d}).

\subsection{Model Lennard-Jones electrolytes in bulk or confined in a slit-like pore}
\label{sec:methods:model}

\textbf{LJ electrolytes.} The model electrolytes consist of neutral and charged LJ particles, representing solvent molecules and ions, respectively, all of which of the same size and same mass $m$ \cite{joly2006liquid}. The electrostatic interactions between ions are screened by the solvent relative permittivity $\varepsilon_s$. All the LJ interactions $U_{LJ}$ were truncated and shifted at a cut-off distance $r_c$.
\begin{equation}\label{lj}
\begin{split}
U_{LJ}(r)=&4\epsilon\bigg[\bigg(\frac{\sigma}{r}\bigg)^{12}-
\bigg(\frac{\sigma}{r}\bigg)^{6}-\bigg(\frac{\sigma}{r_c}\bigg)^{12}+\bigg(\frac{\sigma}{r_c}\bigg)^{6}\bigg],
\end{split}
\end{equation}
assuming the same LJ energy $\epsilon$, and diameter $\sigma$ for interactions between all types of particles.
The cut-off distance for LJ interactions between ions and solvent particles and between solvent particles is $r^*_c=r_c/\sigma=2.5$. Here and in the following, the asterisk represents a quantity in reduced LJ unit. In order to explore the effect of packing due to the presence of explicit solvent particles, we also consider systems where the solvent is only represented by its permittivity, so that only ions are explicitly simulated. In that case, we either use the same LJ interactions between ions with $r^*_c=2.5$ (implicit LJ model) or purely repulsive Weeks-Chandler-Andersen interactions with $r^*_c=2^{1/6}$ (implicit WCA model).

The Coulomb interaction ($U_C$) between LJ (or WCA) ions is:
\begin{equation}
U_{C}(r)=\frac{1}{4\pi\varepsilon_0\varepsilon_s}\frac{q_iq_j}{r}=\frac{\epsilon}{4\pi\varepsilon_0\varepsilon_s}\frac{q^*_iq^*_j}{r^*},
\end{equation}
where $\varepsilon_0$ is the vacuum permittivity and $\varepsilon_s$ the dielectric constant of the solvent, fixed to unity. LJ ions carry either $q^*_i=q_i/\sqrt{4\pi\varepsilon_0\sigma\epsilon}=+1$ or -1, while solvent particles carry no charge. Coulomb interactions are calculated using the particle-particle and particle-mesh (PPPM) method, with a real-space cut-off at $r^*=3.5$.

\textbf{Slit-pore.} The LJ electrolytes are confined in a slit-pore between two atomically flat walls modeled as in Ref.~\citenum{aubouy2003effective}. Each wall consists of 5 layers of a cubic lattice with lattice spacing $\sigma$, and the surface is a two-dimensional square lattice with $n_x\times n_y= 20\times 20$ unit cells, so that the dimensions of the box in the directions parallel to the walls are $L^*_x\times L^*_y= 20\times 20$. The distance $H^*=H/\sigma$ between two inner-most layers (closest to the electrolyte) varies between 12 and 36, ensuring $H\gg\lambda_D$ in all cases studied in this work. There are $n_x\times n_y=N_{w,\text{in}}$ atoms in each layer (hence $10 N_{w,\text{in}}$ atoms for the two walls), whose positions are fixed in space. The total system size, including the walls on both sides, in the $z$ direction is thus  $L^*_z=22,~28,~34$, and 46 for $H^*=12,~18,~24$, and 36, respectively. Periodic boundary conditions are used in the $x$ and $y$ directions only. 
The conversion from LJ to real units requires the knowledge of the diameter of the ions. For example, for a diameter $\sigma=0.5$~nm, a reduced pore width $H^*=12$ corresponds to 6~nm, a reduced bulk ion density $\rho^*_{bulk}=0.0144$ corresponds to 0.0018~nm$^{-3}$, and a reduced surface charge density $\Sigma^*_s=-0.1$ at a reduced temperature $T^*=1$ corresponds to -0.038~$e$/nm$^2$ at 298~K.

Wall atoms interact with the electrolyte via WCA and Coulomb interactions. No LJ interactions between wall atoms were included as they are fixed in space. Only the atoms in the inner-most layer of each wall carry partial charges $q^*_w$ that determine the surface charge density $\Sigma^*_s=N_{w,\text{in}}q^*_w/L^*_xL^*_y$, while all other wall atoms are neutral. We consider $\Sigma^*_s$ varying from -0.05 to -0.5, and accordingly the Gouy-Chapman length $l^*_{GC}$ ($=\frac{\varepsilon_sT^*}{2\pi|\Sigma^*_s|}$) varies from 3.2 to 0.32. With non-zero $q^*_w$, the wall atoms only in the inner-most layers interact with the ions via Coulomb interactions. The Coulomb interactions, including the charged wall atoms and the ionic species, are calculated using the PPPM method, with the same real-space cut-off as for bulk systems, $r^*=3.5$.

\subsection{Non-equilibrium MD/MC with H4D}
\label{sec:methods:h4d}

\begin {figure}[htbp]
\includegraphics [width=3.5in] {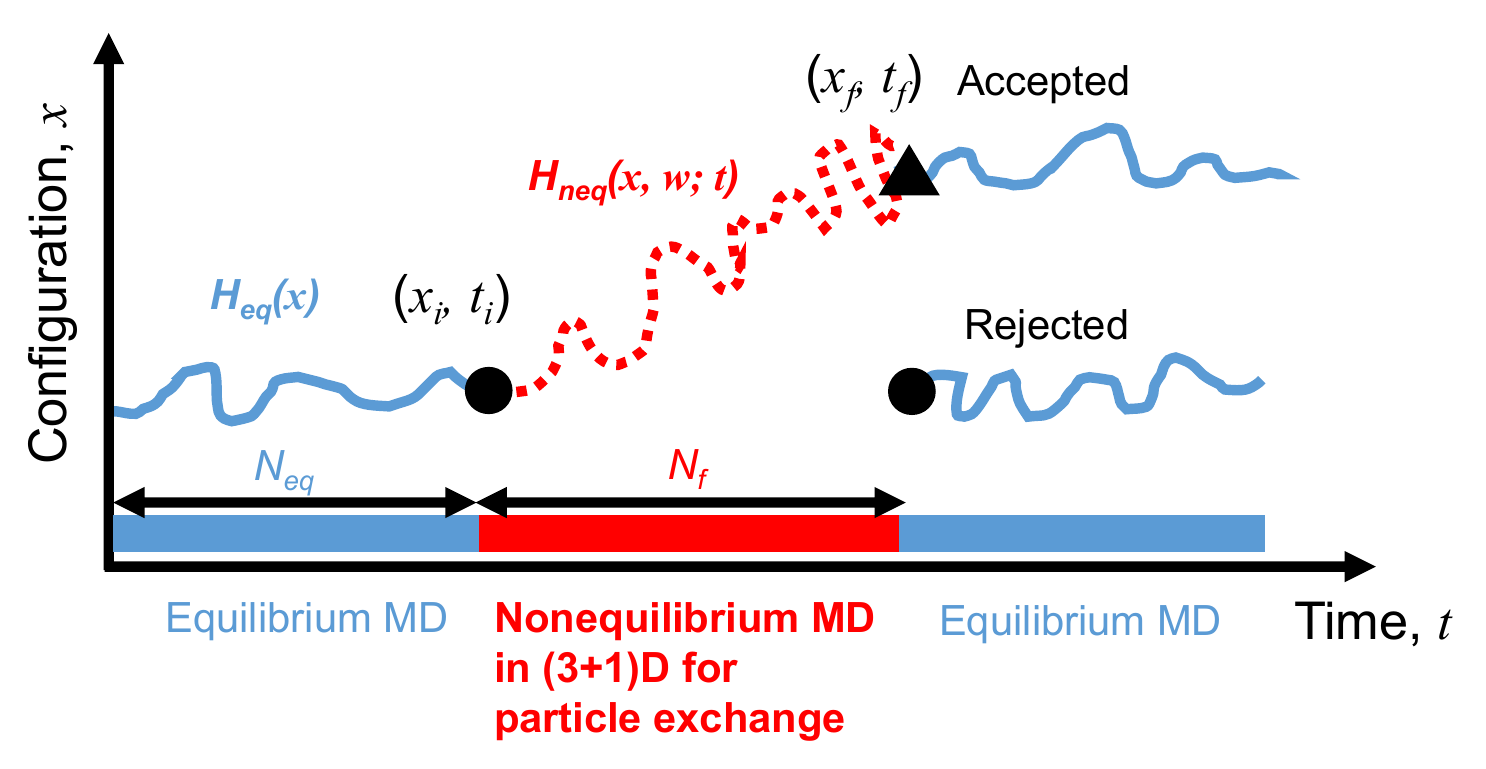}
\centering\caption{
Illustration of a grand-canonical molecular dynamics simulation with H4D: equilibrium MD trajectories with a fixed number of particles (in blue), during which properties are sampled, are interrupted every $N_{eq}$ steps by Monte Carlo trial insertion or deletion moves proposed by nonequilibrium trajectories in the microcanonical ensemble (in red) of $N_f$ steps, during which ``flying'' particles are introduced/removed via a fourth dimension. These trial moves are accepted or rejected according to a Metropolis criterion ensuring that configurations from the equilibrium trajectories sample the Grand-Canonical ensemble, where the chemical potential of the exchanged particles is fixed. The Hamiltonians for the equilibrium and nonequilibrium parts are $H_{eq}$ and  $H_{neq}$, respectively. While the former only depends on the state of the system in 3D (symbolized here by $x$), the latter also depends on a time-dependent "altitude", $w(t)$ (see Eq.~\ref{const-vel}), along the "vertical" direction.
}\label{fig:h4d}
\end{figure}

This section discusses a brief overview of the H4D method, and introduces its extension to confined systems with a bias to alleviate the steric overlaps with confining walls. Monte Carlo simulations in the grand-canonical ensemble are challenging for dense systems, in particular when both steric and electrostatic interactions are present, due to the low acceptance probability of insertion/deletion moves. Hybrid GCMD simulations, whereby during an equilibrium MD simulation Monte Carlo moves are proposed using nonequilibrium MD (NEMD) instead of a "brutal" insertion/deletion, allow to improve the acceptance probability, at the price of increasing the computational cost of each trial move (See Fig.~\ref{fig:h4d}). An elegant type of such trial moves, which proved efficient even for aqueous electrolytes with explicit solvent and ion exchanges, was proposed by Luc Belloni in Ref.~\citenum{belloni2019non} and refined and implemented in the LAMMPS simulation package in Ref.~\citenum{kim2023jcp}. 

\textbf{Basic idea of H4D.} The principle of such hybrid MD/MC simulations is illustrated in Fig.~\ref{fig:h4d}: Equilibrium MD trajectories with a fixed number of particles, during which properties are sampled, are interrupted every $N_{eq}$ steps by Monte Carlo trial insertion or deletion moves proposed by nonequilibrium trajectories in the microcanonical ensemble of $N_f$ steps, during which ``flying'' particles are introduced/removed via a fourth dimension. These trial moves are accepted or rejected according to a Metropolis criterion ensuring that configurations from the equilibrium trajectories sample the Grand-Canonical ensemble, where the chemical potential of the exchanged particles is fixed. The essence of the H4D algorithm is to utilize an auxiliary, non-physical (``vertical'') dimension to facilitate the exchange, which is orthogonal to all other physical dimensions. For three-dimensional systems, during the NEMD all the interactions, including LJ and Coulomb potentials, are treated in (3+1)D, instead of 3D. That is, the potential energy of the system depends not only on the $(x,y,z)$ coordinates of all particles, but also on the ``altitude'' $w$ of the flying particles. In addition, the flying particles are introduced (for insertion) or removed (for deletion) according to a prescribed time-dependent altitude $w(t)$, which is crucial to allow the system to adjust to this perturbation. More details on the computation of the energy in (3+1)D and on the choice of the time-dependent altitude can be found in Ref.~\onlinecite{kim2023jcp}. Here, we use a constant-velocity altitude schedule:
    \begin{equation} \label{const-vel}
        w(t)=v_f\cdot(t-t_i) + w(t_i),
    \end{equation}
where $t\in[t_i,t_f]$ in steps of $\delta t$. The sign of $v_f$ determines whether a trial move is for addition or removal with proper boundary conditions of $w(t)$: For a trial insertion, $w(t_i)=w_{max}$ and $w(t_f)=0$, while for a trial deletion, $w(t_i)=0$ and $w(t_f)=w_{max}$. The key parameters of H4D include the maximum altitude $w_{max}$ and vertical velocity $v_f(t)$, both of which are crucial to determine the efficiency of particle exchange. The altitude schedule (Eq.~\ref{const-vel}) satisfies the detailed balance condition, being symmetric and pre-determined for the trial insertion and deletion moves. In the case of ion-pair exchange, we choose the same boundary conditions of $w(t)$ and the same $v_f$ for both flying ions. In practice, $v_f$ is determined by the number of NEMD steps $N_f$: $v_f=w_{max}\cdot(N_f\delta t)^{-1}$. For an instantaneous (infinitely fast) exchange as in a conventional GC MC, $N_f=0$.

\textbf{Trial exchange moves.} The trial insertion/deletion moves are obtained by propagating the trajectory in the \textit{NVE} ensemble, using a deterministic, time-reversible, and volume-preserving integrator~\cite{mehlig1992hybrid}. In this case, a proposed move is solely determined by the preparation of all momenta; the momenta of flying particles are prepared randomly according to the Maxwell-Boltzmann distribution, and the momenta of non-flying particles are taken from the equilibrium MD. No momentum in the "vertical" direction needs to be assigned, along which an external force determines the altitude of flying particles according to their altitude schedule, and leaves non-flying particles in 3D at zero altitude. Furthermore, we use the symmetric two-end momentum reversal scheme~\cite{chen2014efficient}, which satisfies the detailed balance, since we evolve the systems in the equilibrium phase using MD as well. In this scheme, for every trial addition or deletion, the momenta of all particles are reversed with a probability of one-half both at the beginning and at the end of NEMD.

\textbf{Acceptance probabilities.} A proposed trial move via the H4D method is accepted according to a Metropolis criterion, with a probability ensuring detailed balance~\cite{mehlig1992hybrid, guo2018hybrid, belloni2019non}. Here, we consider electrolytes confined in a slit-like a pore, with impenetrable walls. As commonly done under such circumstances, we introduce a bias in addition to the interactions between the liquid and the wall to favor trial moves consistent with the exclusion of the former from the latter. With this extension of previous work for bulk electrolytes~\cite{belloni2019non, kim2023jcp}, the Metropolis acceptance probability $f_{ins}$ for trial ion-pair insertions ($N_{salt} \rightarrow N_{salt}+1$) is:
\begin{widetext}
\begin{equation}\label{metro_ins_bias}
\begin{split}
    f_{ins}(\vec{r}_a,\vec{r}_c)
    & =\min\bigg[1,
    \exp(-\beta\Delta H_{N_{salt}\rightarrow N_{salt}+1})\exp(\beta\mu_{salt})
    \bigg(\frac{L_xL_yH}{\Lambda_s^{3}}\frac{1}{N_{salt}+1}\bigg)^2
    \frac{B^{del}(\vec{r}_a|\vec{r}_c)}{B^{ins}(\vec{r}_a|\vec{r}_c)}
    \frac{C^{del}(\vec{r}_c)}{C^{ins}(\vec{r}_c)}
    \bigg],\\
    & \equiv \min\bigg[
    1,\exp(-\beta\Delta M+\beta\mu_{salt})
    \bigg],
\end{split}
\end{equation}
\end{widetext}
where $\Delta M$ is the nonequilibrium work, $\Delta H=\Delta U + \Delta K$ the total mechanical energy difference (with $U$ and $K$ the potential and kinetic energies, respectively), $\mu_{salt}$ the chemical potential of the ion pair, and $\Lambda_s  (=\sqrt{\Lambda_+\Lambda_-})$ the geometrical mean of thermal de Broglie wavelengths of an ion pair. Here, $f_{ins}$ depends explicitly on the 3D positions of flying ions ($\vec{r}_a$ for the anion and $\vec{r}_c$ for the cation) via two biases $B(\vec{r}_a|\vec{r}_c)$ and $C(\vec{r}_c)$ to select the cation and anion: the former favors shorter distances between flying ions (as in the bulk~\cite{belloni2019non, kim2023jcp}), while the latter introduces a bias in the position of the flying cation due to the imposed confinement.

{\textbf{Bias for enhanced efficiency.} In our previous work~\cite{kim2023jcp}, $B(\vec{r}_a|\vec{r}_c)$ was found to be crucial to enhance the efficiency of H4D for bulk electrolytes, in addition to modulating the interactions via the fourth dimension. On the one hand, for a trial insertion, the probability to insert the anion at $\vec{r}_a=(x_a,y_a,z_a)$ for a given cation position $\vec{r}_c=(x_c,y_c,z_c)$ is:
\begin{equation}\label{eq:Bins}
B^{ins}(\vec{r}_a,\vec{r}_c)=b^{ins}(x_a|x_c)b^{ins}(y_a|y_c)b^{ins}(z_a|z_c)\cdot V,
\end{equation}
where $V=L_xL_yH$ is the volume of the confined system (with $H$ is the distance between two innermost layers of the confining walls, while $L_x$ and $L_y$ are the simulation box dimensions),
and $b^{ins}$ applies separately to the three directions of space. Without such a bias, $b^{ins}=1/L_i$ for $i\in \{x,y\}$ and $b^{ins}=1/H$ for the $z$ direction so that $B^{ins}=1$  
(see Section~\ref{sec:gcmd} for the functional form and the corresponding parameters used in this work). On the other hand, for a trial deletion, one of the $N_{salt}+1$ anions, knowing the position $\vec{r}_c$ of the flying cation, is chosen with a probability:
\begin{equation}\label{eq:Bdel}
B^{del}(\vec{r}_a,\vec{r}_c)=\frac{B^{ins}(\vec{r}_a,\vec{r}_c)}{\sum_{n=1}^{N_{salt}+1}B^{ins}(\vec{r}_{a,n},\vec{r}_c)}\cdot(N_{salt}+1).
\end{equation}
In the present work, we use the same functional form for the bias in both insertion and deletion trial moves. In the absence of such a bias, $B^{del}=1$.

\textbf{Bias for confined systems.} For confined electrolytes, we introduce another bias, not considered in bulk electrolytes, to in favor of 3D positions of flying ions away from the surfaces. Indeed, a large overlap between a confining wall and a flying ion with its randomly assigned momenta could lead to numerical instabilities or require a quite small $\delta t$ for NEMD, significantly lowering the efficiency of H4D. Even though in principle one could bias the positions of both ions with respect to the walls separately, here we introduce the bias in a sequential manner: first the cation is introduced with the bias due to the confinement on $\vec{r}_c$ with $C(\vec{r}_c)$, then the anion is introduced with the bias for the interionic distance using $B(\vec{r}_a,\vec{r}_c)$. We found this sufficient to ensure that both ions are sufficiently far from the walls.

Here, we discuss the case of confined electrolytes in a slit-pore whose boundaries are at $z=\pm\frac{H}{2}$. The cation is inserted at a position $\vec{r}_c$ with probability:
\begin{equation}\label{eq:Cins}
C^{ins}(\vec{r}_c)= c^{ins}_x(r_x)c^{ins}_y(r_y)c^{ins}_z(r_z)\cdot V,
\end{equation}
where we in fact consider no bias for the periodic directions, \textit{i.e.} $c^{ins}_i(r_i)=L_i^{-1}$ for $i\in\{x,y\}$, and a one-dimensional truncated-shifted Gaussian distribution function along the confinement:
\begin{equation}\label{bias_conf_gauss}
c^{ins}_z(r_z)=\frac{1}{\mathcal{N}_c}\bigg[\exp\bigg(-\alpha_cr_z^2\bigg)-\exp\bigg(\alpha_c\frac{H}{4}^2\bigg)\bigg],
\end{equation}
where $\mathcal{N}_c$ is a normalization constant that ensures $\int_{-H/2}^{H/2}c^{ins}_z(r_z)dr_z=1/H$, and $\alpha_c$ determines the sharpness of the distribution whose mean is at the center of the pore. This choice of bias ensures that the insertion probability vanishes at the confining wall boundaries, \textit{i.e.} $c^{ins}_z(\pm\frac{H}{2})=0$. Without such a bias, $C^{ins}(\vec{r}_c)=1$ for randomly assigned 3D positions as in bulk electrolytes.

For a trial ion-pair deletion, the flying cation is chosen among the $N_{salt}+1$ cations with a probability:
\begin{equation}\label{eq:Cdel}
\begin{split}
C^{del}(\vec{r}_c)
&=\frac{C^{ins}(\vec{r}_c)}{\sum_{n=1}^{N_{salt}+1}C^{ins}(\vec{r}_{c,n})}\cdot(N_{salt}+1)\\
&=\frac{c^{ins}_z(r_z)}{\sum_{n=1}^{N_{salt}+1}c^{ins}_z(r_{z,n})}\cdot(N_{salt}+1).
\end{split}
\end{equation}
The second equality is due to the random choice of the flying-cation position along $x$ and $y$ directions. Again, $C^{del}(\vec{r}_c)=1$ for randomly assigned flying cation 3D positions. The flying anion is then chosen according to Eq.~\ref{eq:Bdel}. We note that such a bias $C(\vec{r}_c)$ can also be applied to the conventional instantaneous MC methods, since it is related to the preparation of the positions of flying ions. 

Once both the cation and anion have been chosen as described above, the probability to accept the trial deletion ($N_{salt}+1 \rightarrow N_{salt}$) is:
\begin{widetext}
\begin{equation}\label{metro_del_bias}
\begin{split}
    f_{des}(\vec{r}_a,\vec{r}_c)
    & = \min\bigg[1,
    \exp(-\beta\Delta H_{N_{salt}+1\rightarrow N_{salt}})\exp(-\beta\mu_{salt})
    \bigg((N_{salt}+1)\frac{\Lambda_s^{3}}{L_xL_yH}\bigg)^2
    \frac{B^{ins}(\vec{r}_a|\vec{r}_c)}{B^{del}(\vec{r}_a|\vec{r}_c)}
    \frac{C^{ins}(\vec{r}_c)}{C^{del}(\vec{r}_c)}
    \bigg]\\
    & \equiv \min\bigg[1,\exp(+\beta\Delta M-\beta\mu_{salt})\bigg].
\end{split}
\end{equation}
\end{widetext}

Similarly, the trial insertion/deletion of solvent particles are generated with the same biases Eqs.~\ref{eq:Cins} and~\ref{eq:Cdel} and accepted with probabilities
\begin{widetext}
\begin{equation}\label{metro_ins_bias_solvent}
\begin{split}
    f_{ins}(\vec{r}_s)
    & =\min\bigg[1,
    \exp(-\beta\Delta H_{N_{solv}\rightarrow N_{solv}+1})\exp(\beta\mu_{solv})
    \bigg(\frac{L_xL_yH}{\Lambda_s^{3}}\frac{1}{N_{solv}+1}\bigg)
    \frac{C^{del}(\vec{r}_s)}{C^{ins}(\vec{r}_s)}
    \bigg],\\
    & \equiv  \min\bigg[
    1,\exp(-\beta\Delta M+\beta\mu_{solv})
    \bigg],
\end{split}
\end{equation}
\end{widetext}
for insertion and 
\newpage 
\begin{widetext}
\begin{equation}\label{metro_del_bias_solvent}
\begin{split}
    f_{des}(\vec{r}_s)
    & = \min\bigg[1,
    \exp(-\beta\Delta H_{N_{solv}+1\rightarrow N_{solv}})\exp(-\beta\mu_{solv})
    \bigg((N_{solv}+1)\frac{\Lambda_s^{3}}{L_xL_yH}\bigg)
    \frac{C^{ins}(\vec{r}_s)}{C^{del}(\vec{r}_s)}
    \bigg]\\
    & \equiv  \min\bigg[1,\exp(+\beta\Delta M-\beta\mu_{solv})\bigg].
\end{split}
\end{equation}
\end{widetext}
for deletion, respectively. Here, $N_{solv}$ is the number of solvent atoms and $\mu_{solv}$ is their chemical potential. In this work, we set $\Lambda_s=\sqrt{\Lambda_+\Lambda_-}=1$ in unit LJ length, and the sign of $\Delta M$ in both trial moves follows the direction of trial insertions as in Refs.~\onlinecite{belloni2019non} and~\onlinecite{kim2023jcp}.

\textbf{Calculation of chemical potentials}
Grand-canonical simulations require the knowledge of the chemical potential of the exchanged species (here the salt, \textit{i.e.} neutral ion pairs, and solvent particles), which is set by the composition and the thermodynamic conditions in the bulk reservoir. Configuration sampling to calculate the chemical potentials was done in the $N_{solv}N_{salt}p^*T^*$ ensemble with $N_{solv}=5000$ and $N_{salt}=100$ at $p^*=p\sigma^3/\epsilon=1$ and $T^*=Tk_B/\epsilon=1$. The desired pressure $p^*=1$ and temperature $T^*=1$ were maintained using the Nos\'e-Hoover barostat and thermostat, with time constants of 1000 and 100~LJ units, respectively. As described in Refs.~\onlinecite{belloni2019non} and~\onlinecite{kim2023jcp}, the chemical potentials were computed using Crooks' fluctuation theorem~\cite{crooks1999entropy} and Bennett's acceptance ratio method~\cite{bennett1976efficient, shirts2003equilibrium}, resulting in: 
$\beta^*\mu^*_{solv}=-1.72\pm0.06$, and $\beta^*\mu^*_{salt}=-11.74\pm0.09$ with $\beta^*=1/T^*$ (see Fig.S1 in SI). For the implicit solvent models, $\beta^*\mu^*_{salt,LJ}=-9.38\pm0.01$ and $\beta^*\mu^*_{salt,WCA}=-8.59\pm0.01$ for LJ and WCA interactions between ions, respectively.

\subsection{GCMD simulations}\label{sec:gcmd}
GCMD simulations were conducted in the $\mu^*_{solv}\mu^*_{salt} A^* H^* T^*$ ensemble, \textit{i.e.} the numbers of both solvent and ion pairs fluctuate in a fixed confined space and their statistics (in particular their average) are determined by their chemical potentials (Fig.~\ref{spm_gcmd_snapshot}). Here, $A^*=L^*_xL^*_y$ and $H^*$ is the distance between the wall atoms in each of the inner-most wall layers.
GCMD simulations are composed of two steps, including equilibrium MD to sample configurations  and non-equilibrium MD/MC steps to exchange solvent or ion pair with reservoir. During equilibrium MD ($N_{eq}$ steps), the equations of motion are integrated using the velocity Verlet algorithm with a timestep $\delta t^*=\delta t\sqrt{\epsilon(m\sigma^2)^{-1}}=0.005$. The desired temperature $T^*=1$ is maintained using the Nos\'e-Hoover thermostat, with a time constant of 100~LJ units. 

During NEMD ($N_{f}$ steps), performed in the $NVE$ ensemble, we also use the velocity Verlet algorithm but with $\delta t^*=0.01$, and the maximum altitude $w^*_{max}$ (see Eq.~\ref{const-vel}) set to unity. While this may not seem the most straightforward choice, we found it beneficial for the acceptance rate of both solvent and ion-pair exchange in bulk electrolytes to use the 3D PPPM estimate of the long-range part of electrostatic interactions (instead of the 4D expression in the presence of flying particles): In addition to its simplicity of implementation, it results in  higher acceptance rates for both solvent and ion-pair exchanges in bulk electrolytes~\cite{kim2023jcp}. The short-range Coulomb contribution was nevertheless calculated in 4D.
Our implementation of the H4D method for confined electrolytes in the LAMMPS simulation package~\cite{plimpton1995fast} is freely available as an update of our previous implementation of the H4D method for bulk electrolytes~\cite{kim2023jcp} (see Data Availability statement). The optimal choice of $N_{eq}$ and $N_f$ is system-specific, and we used $N_{eq}=100$ and $N_f=800$ in most cases.
For each system, all the averaged quantities were computed from six independent trajectories, each of which including 12,000 trial moves of solvent particle or salt-pair exchange with equal probabilities (and following an equilibration including at least 12,000 trial moves).

\begin {figure}[htbp]
\includegraphics [width=3.5in] {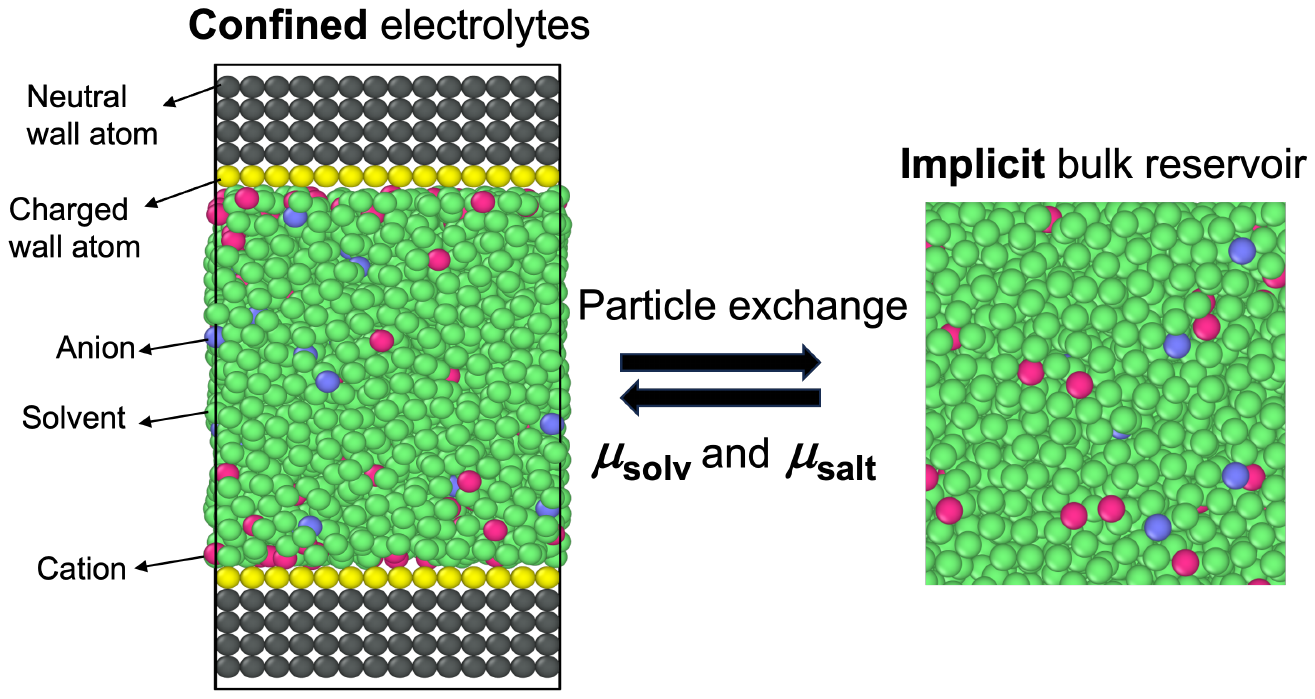}
\caption{Illustration of grand canonical (GC) MD simulation of confined electrolytes with explicit solvent particles in this study. GCMD was conducted using the H4D method, extended to confined electrolytes, in order to enhance the acceptance rate of the salt-pair and solvent exchange trial moves. Chemical potentials used in this study were also computed using H4D: $\beta^*\mu^*_{solv}=-1.72\pm0.06$, and $\beta^*\mu^*_{salt}=-11.74\pm0.09$. See text for details.}
\label{spm_gcmd_snapshot}
\end{figure}

\textbf{NEMD details.}
We apply a bimodal biasing function for the flying-ion distances, as described in Ref.~\citenum{kim2023jcp}. There are several choices for such a biasing function (\textit{e.g.}, a Gaussian distribution with a 3D distance between flying ions). Here, we use a bimodal distribution function for $b^{ins}$, which helps to generate a pair of flying ions that are close to each other yet without large overlap:
\begin{equation}\label{eq:bimodal}
    b^{ins}(x_a|x_c)=
    \sqrt{\frac{\alpha_b}{\pi}}
    \frac{\exp(-\alpha_b(x_{ac}-x_b)^2)+\exp(-\alpha_b(x_{ac}+x_b)^2)}
    {\text{erf}[\sqrt{\alpha_b}(x_b+L_x/2)]-\text{erf}[\sqrt{\alpha_b}(x_b-L_x/2)]}, 
\end{equation}
where $x_{ac}=x_a-x_c$. $x_b$ determines the mean separation of flying ions, and $\alpha_b$ determines the sharpness of the biasing function. Again, $b^{ins}$ is normalized such that $\int_{-L_x/2}^{L_x/2}b^{ins}(x_a|x_c)dx_a=1$. In this work, we used $\alpha^*_b=0.5$ and $x^*_b=2$. The same functional form and parameters are also used in the $y$ and $z$ directions. A small excluded volume was also applied: $V^*_{ex}=0.125$ for early rejection (See details in Ref.~\citenum{kim2023jcp}). For the bias in the confined systems (see Eq.~\ref{bias_conf_gauss}), $\alpha^*_c$ varies from 0.5 to 6 with decreasing separation distance. In order to avoid numerical instabilities, one can increase $\alpha^*_c$ to obtain a narrow Gaussian distribution near the center of the slit pore.

Since for the moderate salt concentrations considered in the present work,  with an implicit solvent the packing fraction is low, the potential gain in efficiency with the H4D method with respect to conventional GCMC (corresponding $N_f=0$) is not worth the additional computational cost. Therefore, for implicit solvent models, trial insertions and deletions of the ions are carried out instantaneously (this corresponds to an infinite vertical velocity, with $N_f=0$) as in conventional GCMC. We can illustrate the benefit of the H4D method for the explicit solvent case on the system with pore size $H^*=12$ and surface charge density $\Sigma^*_s=-0.1$, for which the H4D with $N_f=800$ and $N_{eq}=100$ achieves acceptance rates $P_{acc}$ of $0.28\pm0.05$ and $0.094\pm0.001$ for solvent particle and salt-pair exchange, respectively. For comparison, we performed 3 independent runs, each with 6,000 conventional GCMC trial moves, of the same system (with $N_f=0$ and $N_{eq}=100$), resulting in $P_{acc}=0.0028\pm0.0005$ for the solvent exchange and $P_{acc}=0$ for the salt-pair exchange (not a single trial was accepted). In terms of the efficiency metric, $E^t_f=P_{acc}/(N_{eq}+N_f)$, this means a gain with H4D larger than one order of magnitude for the solvent exchange -- and a gain that cannot even be quantified for salt-pair exchange since the efficiency is null with GCMC. We note that in our previous work for bulk dilute electrolytes~\cite{kim2023jcp}, a 10$^4$ gain in efficiency for salt-pair exchange could be achieved by fine-tuning all the parameters (which was not done here), which suggests that there is room for further improvement in the present case.

\subsection{Computation of observables}
\label{sec:methods:observables}

We analyze the composition and structure of the confined liquid by computing the ionic and solvent density profiles, $\rho^*_\pm(z^*)$ and $\rho^*_{\text{solv}}(z)$, using histograms with bin width $\Delta z^*=0.1$, and symmetrized with respect to the pore center ($z^*=0$). Since only neutral ion pairs or solvent particles are exchanged in GCMD, the electroneutrality constraint
\begin{equation}
\int_{-H^*/2}^{H^*/2}dz^*\rho^*_q(z^*)=-2\Sigma^*_s,
\end{equation}
with $\rho^*_q(z^*)=\rho^*_+(z^*)-\rho^*_-(z^*)$ the charge density, is always satisfied. We also compute the cumulative mean-charge density from the charged surface
\begin{equation}\label{cumulative_Q}
Q^*(z^*)=\Sigma^*_s+\int_{-H^*/2}^{z^*}\rho^*_q\big(z^*\big)dz^*,
\end{equation}
which decays from $\Sigma^*_s$ at $z^*=-H^*/2$ to 0 at $z^*=0$ (the centre of the slit-pore), if $H^*\gg\lambda^*_D$. 

The decay of $Q^*(z^*)$ also provides an estimate of the thickness of the electrical double layer at the surface. Due to the short-range repulsion between the ions and the wall, there is a region from which ions are excluded where $\rho^*_q=0$ and $Q^*(z^*)=\Sigma^*_s$.
According to the linearized PB equation, beyond this exclusion region, for dilute electrolytes with weakly charged walls we expect an exponential decay as
\begin{equation}\label{dhq}
Q^*(z^*_d)\approx \Sigma^*_s\exp[-(z^*_d-s^*)/\lambda^*_{D}],
\end{equation}
where $z^*_d$ is the distance from the charged wall, and $s^*$ is an effective width of the exclusion layer. We note again that this approximation assumes $H^*\gg\lambda^*_D$, which is satisfied in all the cases considered in the present work. As explained in Section~\ref{sec:donnan:effectivecharge}, for larger $\Sigma^*_s$ and sufficiently large distances from the surface, one can approximate the decay using an effective surface charge density $\Sigma^*_{\text{eff}}=a_{\text{eff}}\Sigma^*_s$:
\begin{equation}\label{mdhq}
Q^*(z^*_d)\approx \Sigma^*_{\text{eff}}\exp[-(z^*_d-s^*)/\lambda^*_{D}]
\end{equation}
with $\Sigma^*_{\text{eff}}=a_{\text{eff}}\Sigma^*_s$. Similar expressions are obtained for the corresponding potential and ionic density profiles. As a result of the exclusion of ions from both walls due to their finite size, the effective width of the slit-pore is not the bare distance $H^*$ between the inner-most layers of the confining walls, but an effective size $H^*_{\text{eff}}=H^*-2s^*$. Based on the simulation data (see Section~\ref{sec:results}), in the following we estimate $H^*_{\text{eff}}$ using  $s^*=0.9$ whenever comparing the simulation results to analytical predictions involving the pore size.

\section{Results and discussion}\label{sec:results}

In this section, we discuss the results of GCMD simulations for the LJ electrolytes with explicit solvent particles confined in a slit-like pore with negatively charge surfaces. The system is in an equilibrium with a reservoir setting the chemical potentials of the salt and of solvent particles, whose values are computed as described in Section~\ref{sec:methods:h4d}. The salt concentration in this reservoir ($\rho^*_{bulk}$=0.0144) is sufficiently small for Debye-H\"uckel theory to apply, and the corresponding Debye screening length is $\lambda^*_D=1.66$. As shown in Fig. S2 of the SI, we observe in the bulk simulations that the charge-charge correlation function indeed decays exponentially with the expected decay length. 

\begin {figure*}[htbp]
\includegraphics [width=6.5in] {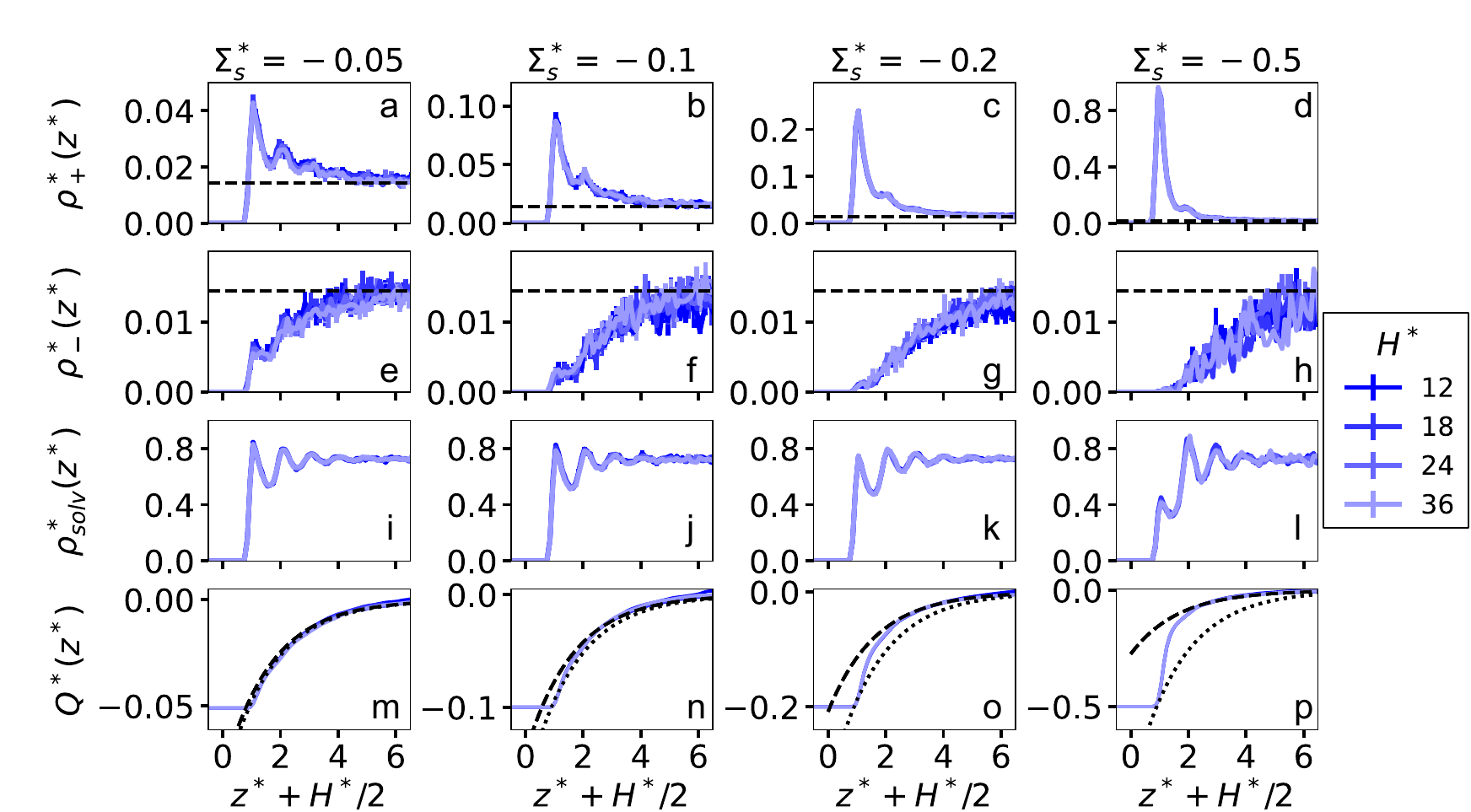}
\caption{
Local densities of cations $\rho^*_+$ (a-d), anions $\rho^*_-$ (e-h), and solvent $\rho^*_{\text{solv}}$ (i-l) confined in charged slit pores of different widths $H^*$ and surface charge densities $\Sigma^*_s$, and the corresponding cumulative mean-charge density $Q^*(z^*)$ (m-p) defined by Eq.~\ref{cumulative_Q}. Colors represent different $H^*$, as indicated in the legend on the right. Horizontal dashed lines in $\rho^*_+$ and $\rho^*_-$ represent the bulk salt concentration $\rho^*_{\text{salt}}$ in the reservoir. In panels (m-p), the dotted and dashed lines are the predictions from linearized PB theory with the bare and effective surface charge densities, given by Eqs.~\ref{dhq} and~\ref{mdhq}, respectively.
}
\label{fig:localdensity}
\end{figure*}

Figure~\ref{fig:localdensity} displays the local densities of the cations (panels a-d), anions (panels e-h), and solvent particles (panels i-l) confined in slit-pores with different sizes ($H^*=12,~18,~24,~\text{and}~36$) and surface charge densities ($\Sigma^*_s=-0.05,~-0.1,~-0.2,~\text{and}~-0.5$). 
Each column displays the profiles for the same $\Sigma^*$, with the positions $z^*$ shifted by $H^*/2$ to align the positions of the ``left'' walls (at $z^*=-H^*/2$), and the colors in each panel correspond to the various pore sizes. For all the considered $\Sigma^*_s$, the shifted profiles for the various $H^*$ overlap. Deviations from the bulk behaviors emerge only at the surfaces under these mild confinement conditions, for which $H^* \gg \lambda^*_D, l_{GC}$. The cumulative charge distribution, computed from Eq.~\ref{cumulative_Q}, is also reported in Figure~\ref{fig:localdensity} (panels m-p).

\subsection{Weakly charged pores}

The leftmost column of Fig.~\ref{fig:localdensity} (a,e,i,m) displays the results of dilute LJ electrolytes confined in a weakly charged pore ($\Sigma^*_s=-0.05$ and $l^*_{GC}=3.2$). As expected, electrical double layers form at the charged surfaces, with an excess of counterions (panel a) and a depletion of co-ions (panel e) near the walls, with respect to the salt concentration far from them. In addition, the ionic density profiles display oscillations near the wall due to packing effects dominated by the solvent density (panel i), which is much larger than that of the ions and displays at least three well defined layers. This is in stark contrast to implicit solvent models where no such layering is observed in the ion density profiles (see Section~\ref{sec:results:implicit}). The cumulative mean-charge distribution $Q^*(z^*)$ (panel m) follows an exponential decay with characteristic length $\lambda^*_D$, well described by the linearized PB result (Eq.~\ref{dhq}), to zero at the center of the pore. In this case the difference between the bare and renormalized surface charge densities is small ($a_{\text{eff}}\approx 0.94$). Finally, we note that $Q^*(z^*)$ does not display any oscillations despite that of the ionic density profiles. Such an observation is likely due to the simplicity of the solvent model, which does not include an explicit charge distribution (dipolar or more complex) which would contribute to $Q^*(z^*)$~\cite{PhysRevLett.107.166102,PhysRevLett.117.048001,gonella2021water,nickel2024water,schlaich2024water}.

\subsection{Strongly charged pores}
With more negative $\Sigma^*_s$ (and decreasing $l^*_{GC}$), the structure of the EDLs evolves with further enrichment of cations (Figs.~\ref{fig:localdensity}a-d) and depletion of anions (Figs.~\ref{fig:localdensity}e-h) near the wall. Due to the finite size of the ions and solvent molecules, the increasing density of cations leads to the progressive depletion of solvent particles at the interface (Figs.~\ref{fig:localdensity}i-l). For $\Sigma^*_s=-0.5$, the cation density is even larger than that of the solvent in the first fluid layer in contact with the surface.
Since the solvent density near the wall is depleted (even more so that $|\Sigma^*_s|$ is large) due to the increased density of counterions, so is the average solvent density inside the pore (See Fig. S3 in the SI). 

For large surface charge densities, the decay of $Q^*(z^*)$ (Figure~\ref{fig:localdensity}m-p) departs significantly from the Debye-H\"uckel result (Eq.~\ref{dhq}). Close to the wall, it decays much faster due to the large concentration of ions in the first layer, consistently with the Gouy-Chapman-Stern picture of the interface. Beyond this layer, the decay is well described by the linearized PB result with a renormalized surface charge density (Eq.~\ref{mdhq} and the dashed lines in Figure~\ref{fig:localdensity}o-p). The failure from the bare DH prediction is not surprising, since this is a regime where ion-wall interactions are large compared to ion-ion interactions ($l^*_{GC}\lesssim\lambda^*_D$)~\cite{markovich2021charged,herrero2021Poisson}. Importantly, such an agreement suggests that (i) the effective charge describing the far-field is well predicted by Eq.~\ref{eq:aeff}, and (ii) the decay length at long distance is well described by $\lambda^*_D$.

\subsection{Donnan equilibrium}
\begin {figure}[htbp]
\includegraphics [width=2in] {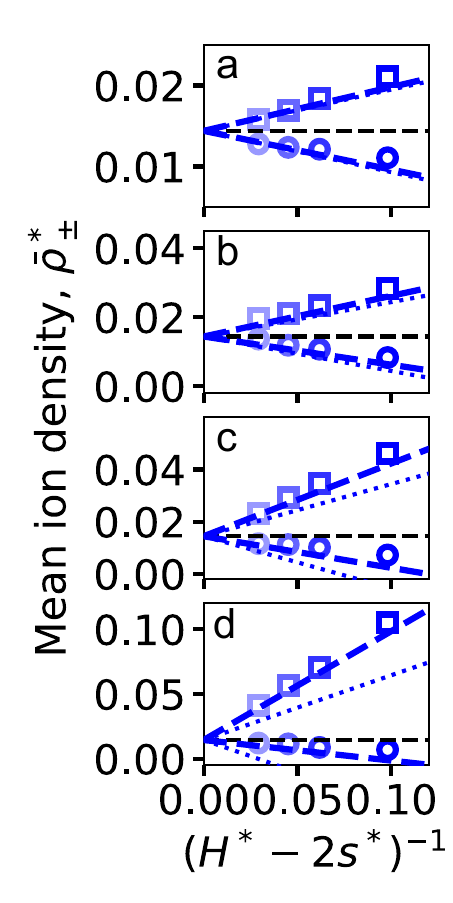}
\caption{
Mean cation ($\bar{\rho}^*_{+}$, squares) and anion ($\bar{\rho}^*_{-}$, circles) densities in slit pores with different widths $H^*$ and different surface charge densities: $\Sigma^*_s$ = -0.05 (a), -0.1 (b), -0.2 (c), and -0.5 (d). Results are plotted as a function of the inverse of the effective width of the pore, $H^*_{\text{eff}}=H^*-2s^*$, where the parameter $s^*=0.9$ accounts for the excluded volume between the liquid and the wall atoms (see Section~\ref{sec:methods:observables}). The dotted and dashed blue lines are the predictions from linearized PB theory with the bare (Eqs.~\ref{lpb+} and~\ref{lpb-}) and renormalized (Eqs.~\ref{mlpb+} and~\ref{mlpb-}) surface charge densities, respectively. The horizontal black dashed line represents the bulk ion density $\rho^*_{\text{salt}}$ in the reservoir.
}
\label{fig:donnan}
\end{figure}

The different composition of the confined solution with respect to the bulk reservoir, which characterizes the Donnan equilibrium, can be quantified by the mean cation and anion densities insides the pore. Since the linearized PB theory predicts linear scalings as a function of the inverse of the distance between the walls (see Eqs.~\ref{lpb+} and~\ref{lpb-}), the results are reported in Fig.~\ref{fig:donnan} as a function of the inverse of the effective pore width, $H^*_{\text{eff}}=H^*-2s^*$, where the parameter $s^*=0.9$ accounts for the excluded volume between the liquid and the wall atoms (see Section~\ref{sec:methods:observables}). The predictions (Eqs.~\ref{lpb+} and~\ref{lpb-}) are shown as dotted lines. They are accurate for the smallest considered surface charge densities, but as $|\Sigma^*_s|$ increases, they underestimate both the cation and anion mean densities. Nevertheless, even for the larger values considered, the scaling remains linear in $1/H^*_{\text{eff}}$ (which validates its choice of the relevant definition of the pore width as well as the corresponding value of $s^*$) and consistent with the limit $\rho^*_\pm\to \rho^*_{salt}$ for very large pores (where interfacial effects become negligible). In fact, all results are very well described by the predictions of linearized PB with the effective charge $\Sigma^*_{\text{eff}}=a_{\text{eff}}\Sigma^*_s$ (see Eqs.~\ref{mlpb+} and~\ref{mlpb-}), shown as dashed blue lines. This good agreement reflects the better accuracy of the non-linear PB theory, whose result is implicitly taken into account via the renormalized surface charge density, at least in the present regime where $\lambda^*_D$ is small compared to $H^*_{\text{eff}}$.

\subsection{Comparison to implicit solvent models}\label{sec:results:implicit}

\begin {figure*}[htbp]
\includegraphics [width=7in] {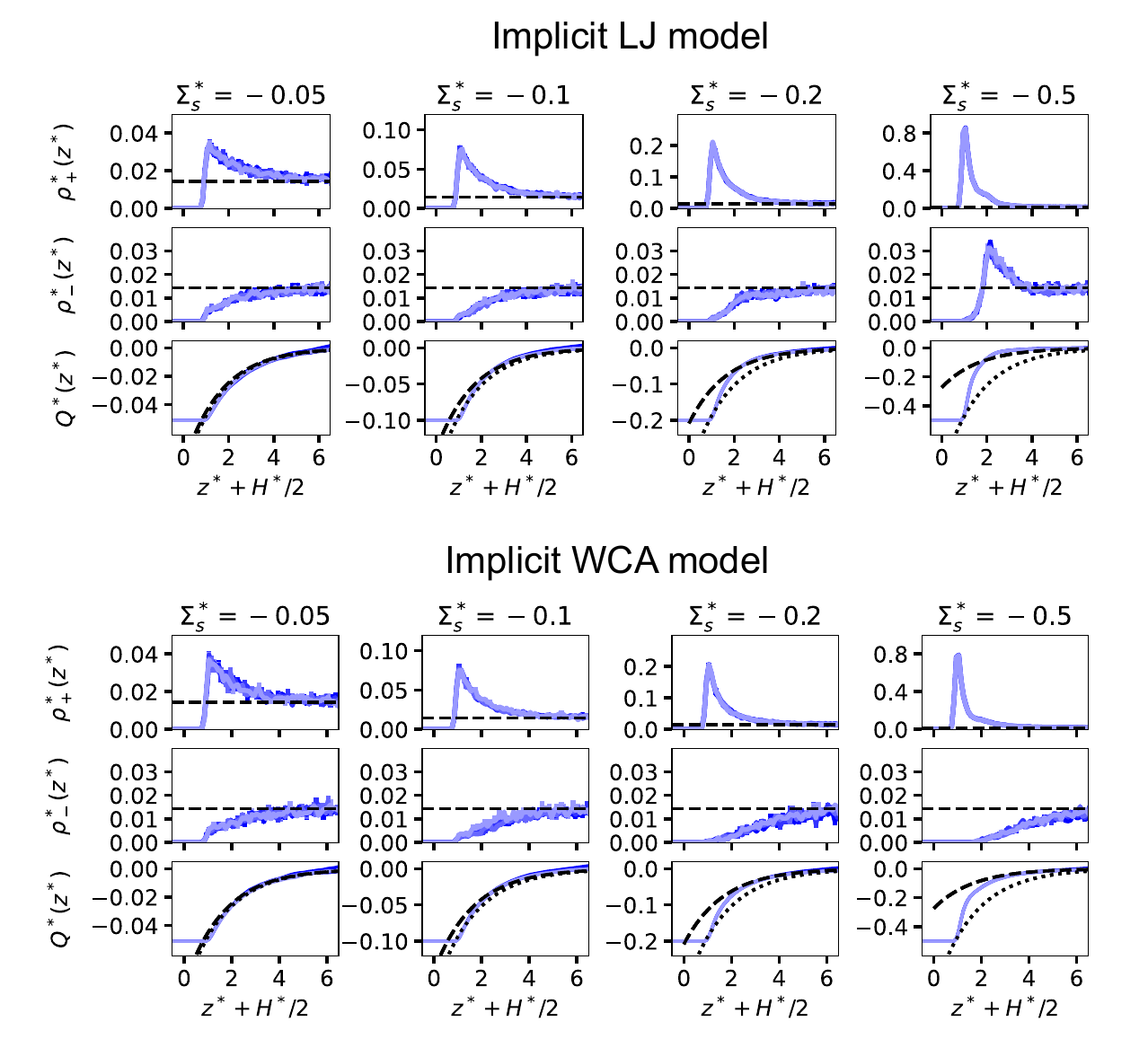}
\caption{
Local densities of cations $\rho^*_+$ and anions $\rho^*_-$ for the two implicit-solvent electrolyte models (implicit LJ, top, and WCA, bottom; see Section~\ref{sec:methods:model}) confined in a slit of different width $H^*$ and surface charge densities $\Sigma^*_s$, and the corresponding cumulative mean-charge density $Q^*(z^*)$ defined by Eq.~\ref{cumulative_Q}. Colors represent different $H^*$, as indicated in the legend on the right of Fig.~\ref{fig:localdensity}. Horizontal dashed lines in $\rho^*_+$ and $\rho^*_-$ represent the bulk salt concentration $\rho^*_{\text{salt}}$ in the reservoir, while for $Q^*$, the dotted and dashed lines are the predictions from linearized PB theory with the bare and effective surface charge densities, given by Eqs.~\ref{dhq} and~\ref{mdhq}, respectively.
}
\label{fig:local_density_implicit}
\end{figure*}

In order to investigate the solvent packing effect, we conducted additional GCMD simulation of two dilute electrolytes with explicit ions (with short-range LJ or WCA interactions, see Section~\ref{sec:methods:model}) and no explicit solvent particles (only a permittivity screening the Coulomb interactions between ions, as in the model with explicit solvent considered here). For bulk electrolytes, all three models predict long-range screening behaviors well described by Debye-H\"uckel theory, despite the differences in short-range structures in ion-ion radial distribution functions (See Fig.~S2 in SI). 

\textbf{Interfacial structure.}
In both implicit cases, local ion densities inside the pores (Fig.~\ref{fig:local_density_implicit}) show no oscillations at all $\Sigma^*_s$'s, which further supports the conclusion that the oscillations with an explicit solvent in Fig.~\ref{fig:localdensity} originate from the solvent packing effect. For the three smaller densities considered ($|\Sigma^*_s|\leq 0.2$), the simulation results are very similar to each other and to the explicit-solvent case, except for the above-mentioned oscillations. In particular, the cumulative charge density is well described by the linearized PB prediction with the bare charge for the lowest $|\Sigma^*_s|$, and at long distance with the renormalized surface charge for intermediate $|\Sigma^*_s|$ (but not at short distance).

The main difference between the two implicit models is observed for $\Sigma^*_s=-0.5$. While the predictions of the implicit WCA model are very close to the explicit solvent case, the implicit LJ model predicts a peak in the anion concentration close to the cations adsorbed at the surface (at a distance from the surface corresponding approximately to the shoulder of $\rho^*_+$ next to the first peak). Such an excess of anions close to the adsorbed cations results from the short-range attraction between ions, which is absent in the implicit WCA case. 
In our explicit solvent model, the short-range attraction between cations and solvent particles is identical to that between cations, so that the electrostatic energy cost of increasing the cation density is not mitigated by the replacement of solvent molecules by cations next to the surface. In contrast, the implicit LJ model includes a short-range attraction which mitigates the repulsion between like-charged ions and increases the attraction between ions oppositely-charged ions. This is for example known to shift the bulk phase diagram (see \textit{e.g.} Refs.~\citenum{de1990melting,ahmed2009phase}).
Since such ionic correlations, of course, are not included in the PB theory, the latter is insufficient to predict the cumulative charge, even at long distances since the presence of anions close to the surface also modifies the electric field experienced by ions far from it.

\textbf{Excess ion density.} The structure of EDL is closely related to the Donnan equilibrium between the charged pore and the bulk reservoir. The predictions of the three electrolyte models for the excess ion density $\Delta\rho^{*}_{\text{ex}}$ with respect to the bulk reservoir (see Eq.~\ref{eq:excess_mean_eff}) are reported with different colors in Fig.~\ref{fig:donnan_excess}, as a function of the inverse of the effective width of the pore, $H^*_{\text{eff}}=H^*-2s^*$, for all the considered surface charge densities, from $\Sigma^*_s=-0.05$ (circles) to $\Sigma^*_s=-0.5$ (squares) (The mean cation and anion densities for the implicit solvent models can be found in Fig.~S4 in SI). For weakly charged surfaces ($l^*_{GC}\gtrsim\lambda^*_D$), $a_{\text{eff}}\approx1$ and the excess ion density is small ($\Delta\rho^{*}_{\text{ex}}\approx0$). For larger $|\Sigma^*_s|$, both implicit models follow the linear trend observed with the explicit solvent and predicted by the linearized PB theory using a renormalized charge (see Eq.~\ref{eq:excess_mean_eff}), also shown as dashed lines. However, while the implicit WCA model closely follows the predictions with the explicit solvent, the implicit LJ model consistently overestimates the salt concentration, even more so that $|\Sigma^*_s|$ is large. 

\begin {figure}[htbp]
\includegraphics [width=3in] {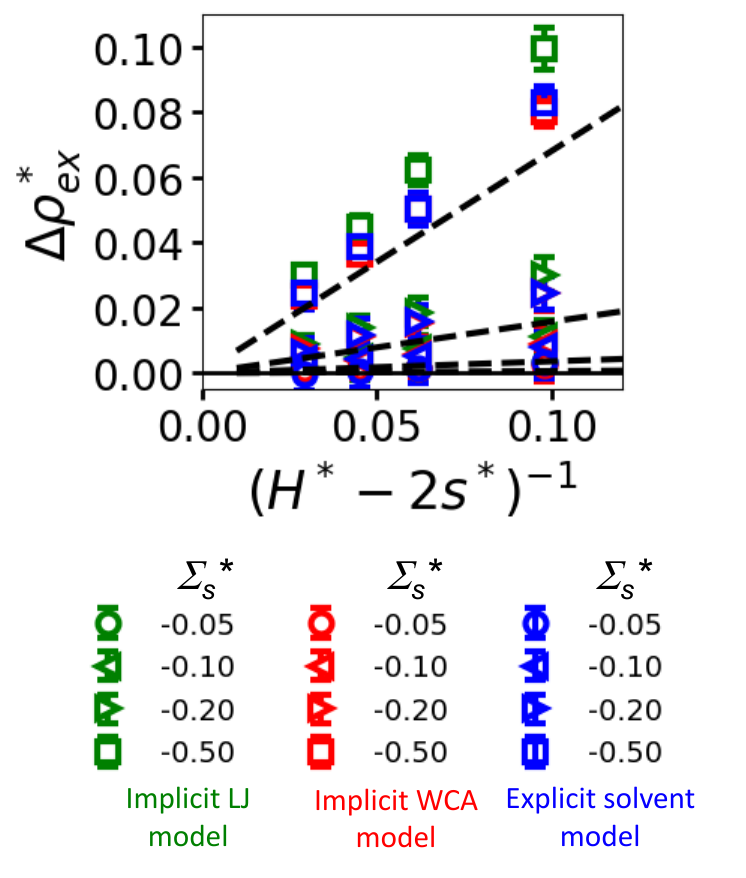}
\caption{Mean excess ion density $\Delta\rho^{*}_{\text{ex}}$ of three electrolytes models: Blue for the explicit solvent model, red for the implicit WCA model, and green for the implicit LJ model. Four different markers represent four different surface charge densities $\Sigma^*_s$. Results are plotted as a function of the inverse of the effective width of the pore, $H^*_{\text{eff}}=H^*-2s^*$, where the parameter $s^*=0.9$ accounts for the excluded volume between the liquid and the wall atoms (see Section~\ref{sec:methods:observables}). The black dashed lines represent the prediction (Eq.~\ref{eq:excess_mean_eff}) of the linearized PB theory with a renormalized surface charge density $a_{\text{eff}}\Sigma^*_s$ (see Eq.~\ref{eq:aeff}).
}
\label{fig:donnan_excess}
\end{figure}

In order to better quantify the differences between the three electrolyte models, we fit the simulation results of Fig.~\ref{fig:donnan_excess} by Eq.~\ref{eq:excess_mean_eff} treating the factor $a_{\text{eff}}$ as a fitting parameter.
The results are summarized in Table~\ref{tab:aeff}, together with the analytical predictions Eq.~\ref{eq:aeff}, or the values resulting from fitting to Eq.~\ref{eq:excess_mean_eff} the excess salt density obtained by numerically integrating the ionic density profiles of the ions in the thin EDL approximation, where the potential entering in their Boltzmann distribution is given by Eq.~\ref{eq:dhb}. In the latter case, exploiting the symmetry with respect to the center of the pore, this reads explicitly:
    \begin{equation}
    \Delta\rho^{*}_{\text{ex}}(H^*)= \frac{2}{H^*}\int_{0}^{{H^*}/2}\rho^{*}_{\text{salt}}[e^{-\Psi^*(z^*_d)}+e^{+\Psi^*(z^*_d)}]\, {\rm d}z_d-2\rho^{*}_{\text{salt}}
    \end{equation}
with $\Psi^*(z^*_d)$ from Eq.~\ref{eq:dhb}.
Consistently with Fig.~\ref{fig:donnan_excess}, these results indicate that Eq.~\ref{eq:aeff} is only accurate for sufficiently small surface charge density, and overestimates $a_{\text{eff}}$ (hence underestimates $\Delta\rho^{*}_{\text{ex}}$) for larger $|\Sigma^*_s|$. In contrast, the non-linear thin-EDL approximation is accurate for all the considered cases (for which $\lambda^*_D$ is smaller than $H^*$, as already noted). This suggests that the limitations of Eq.~\ref{eq:aeff} are not due to the breakdown of the mean-field description, but rather to the charge renormalization approximation, which is not sufficient to capture the overall excess ion density because it focuses on the behavior far from the walls. 

\begin{table}
\centering
    \begin{tabular}{ |c||c|c|c|c|c| } 
     \hline
     $\Sigma^*_s$  & $a_{\text{eff}}$  &  $a_{\text{eff}}$  &  $a_{\text{eff,fit}}$  & $a_{\text{eff,fit}}$ & $a_{\text{eff,fit}}$ \\ 
      & (Eq.~\ref{eq:aeff}) & (Eq.~\ref{eq:dhb}) & (explicit) & (implicit WCA) & (implicit LJ)\\ 
     \hline
     \hline
     -0.05 & 0.94 & 0.75 &  0.73 & 0.99 & 0.95 \\ 
     -0.1 & 0.82 & 0.58 & 0.57 & 0.66 & 0.55 \\ 
     -0.2 & 0.61 & 0.38 & 0.37 & 0.44 & 0.30 \\ 
     -0.5 & 0.31 & 0.18 & 0.16 & 0.21 & 0.015 \\
     \hline
    \end{tabular}
\caption{
Charge renormalization factor $a_{\text{eff}}$ predicted by Eqs.~\ref{eq:aeff} and~\ref{eq:dhb} (see text for detail) and by numerically fitting the simulation results for the excess ion density $\Delta\rho^{*}_{\text{ex}}$ for the three electrolyte models (see Fig.~\ref{fig:donnan_excess}) to Eq.~\ref{eq:excess_mean_eff}. 
}
\label{tab:aeff}
\end{table}

Table~\ref{tab:aeff} also confirms that the implicit WCA model overall provides a better agreement with the explicit solvent than the implicit LJ model. The most significant difference is observed for the largest surface charge density, $\Sigma^*_s=-0.5$, where the implicit LJ model predicts $a_{\text{eff}}\approx0$, \textit{i.e.} a perfect screening of the surface charge by the nearby ions, which also include co-ions in that case (see Fig.~\ref{fig:local_density_implicit}). We further note that when the field due to the surface charge is completely screened, the salt concentration inside the pore is equal to that of the reservoir (\textit{i.e.} no Donnan exclusion). Finally, the good agreement between the explicit solvent and implicit WCA model indicates that packing effects do not play a significant role in the Donnan exclusion in the considered cases (even though it does lead to oscillations in the ionic density profiles, as discussed above).

\section{Conclusion}
\label{sec:conclusion}

We investigated the Donnan equilibrium of coarse-grained dilute electrolytes confined in charged slit-pores in equilibrium with a reservoir of ions and solvent. This was achieved using an extension of a recently developed hybrid Grand canonical / nonequilibrium molecular dynamics simulation method (\textit{J. Chem. Phys.} 151, 021101 (2019) and \textit{J. Chem. Phys.} 159, 144802 (2023)) to confined systems, which enhances the efficiency of solvent and ion-pair exchange, using insertion/deletion via a fourth spatial dimension. We restricted ourselves to the case of dilute electrolytes in the thin electric double-layer limit and explored the influences of the pore size and of the surface charge density, as well as packing effects due to an explicit solvent by comparison with implicit solvent models, on the co- and counter-ion distributions inside the pore, the cumulative charge density profiles at the interface, and the mean ion concentrations of the confined liquid, which differ from that in the electrolyte reservoir. 

We showed that the validity range of linearized Poisson-Boltzmann theory to predict the Donnan equilibrium of dilute electrolytes can be extended to highly charged pores, by considering renormalized surface charge densities, which can be computed analytically by comparing the full and linearized PB equations, instead of the bare ones. By comparing with simulations of implicit solvent models of electrolytes, we find that for the small salt concentration considered here, an explicit solvent introduces oscillations in the ionic density profiles, but has a limited effect on the excess salt concentration inside the pore. We note that the ability of implicit solvent models to predict ``macroscopic'' quantities such as the excess salt concentration, or the ionic density profiles far from the walls, arises from a cancellation of errors between over/underestimates near the walls due to the absence of oscillations in the ionic profiles, even for small surface charge densities (see Figs.~\ref{fig:localdensity} and~\ref{fig:local_density_implicit}), as already observed in molecular simulations of closed systems with an explicit solvent (see \emph{e.g.} Refs.~\citenum{marry2003equilibrium, tournassat2009comparison}). Such a cancellation of errors might not be as favorable for other properties, in particular dynamical ones. In the low concentration and thin electric double-layer limit considered here, the main limitations of the analytical predictions are not due to the breakdown of the mean-field description, but rather to the charge renormalization approximation, which is not sufficient to capture the overall excess ion density because it focuses on the behavior far from the walls. 

In the models of electrolytes considered in the present work, the dielectric response of the solvent is taken into account only by its permittivity to screen the electrostatic interactions between ions.
While it has the advantage of allowing us to disentangle the effects of long-range electrostatic interactions from short-range van der Waals interactions, this is a very crude representation since the charge distribution within solvent molecules results in more complex interactions between themselves, with ions and with charged walls. Furthermore, the computation of electrostatic interactions in the confined case assumes that there is no dielectric contrast between the liquid and the wall, which is the exception rather than the rule.

In the present work, the charge was uniformly distributed on the solid surface. However, surface charge heterogeneities have important effects on the properties of interfacial electrolytes, for example on the distribution of ions and their solvation at the surface~\cite{marry2008structure,tournassat2009comparison}, as well as the solid-liquid friction and electrokinetic response~\cite{botan2011hydrodynamics,rotenberg2013electrokinetics,simonnin2018mineral,xie2020liquid,mangaud2022chemisorbed}. In addition, longer-range effects may arise when surface charge heterogeneities occur on scales comparable to the electrostatic correlation length in the electrolyte (the Debye screening length at sufficient small concentration), as discussed \textit{e.g.} in Refs.~\citenum{ben2013interaction,mussotter2020heterogeneous,bier2022structure}. We leave the study of surface charge heterogeneity on the Donnan equilibrium for future work.

The present Grand canonical simulation using the H4D can also be applied with an explicit solvent such as the SPC/E water model~\cite{belloni2019non,kim2023jcp}, as well as more concentrated electrolytes than the one considered here, or multivalent ions. This would provide insights into the effects of ion-ion and ion-solvent correlations on the properties of the confined electrolytes, and to explore the possible role of the Donnan equilibrium in the observed scaling of Surface Force Balance measurements with the concentration~\cite{smith2016electrostatic,lee_underscreening_2017,elliott_known-unknowns_2024} or in the breakdown of electroneutrality in nanopores~\cite{boda2006effect,levy2020breakdown,green2021conditions}.
In addition, the present extension of the H4D method to slit pores could be further extended to more complex geometries, including disordered porous materials such as ion exchange membranes.
\section*{Acknowledgements}
We thank Luc Belloni for fruitful discussions. This project received funding from the European Research Council under the European Union’s Horizon 2020 research and innovation program (grant agreement no. 863473).

\section*{Data Availability Statement}
Our implementation of the H4D method in the LAMMPS simulation package is freely available at \href{https://github.com/Jeongmin0658/h4d_lammps}{https://github.com/Jeongmin0658/h4d\_lammps}.
All the data presented in this work will be provided upon reasonable requests.

\bibliography{manuscript}
\end{document}